\documentclass[12pt]{article}

\usepackage{amsmath,amsfonts,enumerate,array}
\usepackage{graphicx}
\usepackage[usenames]{color}
\usepackage[english]{babel}
\usepackage{fancybox}
\usepackage{chngpage} 

\usepackage{tikz-cd}

\newcommand{\captionfonts}{\small}

\makeatletter  
\long\def\@makecaption#1#2{%
  \vskip\abovecaptionskip
  \sbox\@tempboxa{{\captionfonts #1: #2}}%
 \ifdim \wd\@tempboxa >\hsize
    {\captionfonts #1: #2\par}
  \else
    \hbox to\hsize{\hfil\box\@tempboxa\hfil}%
  \fi
  \vskip\belowcaptionskip}
\makeatother   

\usepackage{mciteplus}

\usepackage[colorlinks=true,      linkcolor=blue,      urlcolor=blue,      filecolor=blue,      citecolor=blue,
      pdfstartview=FitH,     pdfpagemode=UseNone,      bookmarksopen=true]{hyperref}  
      
\usepackage[all]{hypcap}     

\usepackage{cite}

\begin{document}

\numberwithin{equation}{section}


\mathchardef\mhyphen="2D


\newcommand{\be}{\begin{equation}} 
\newcommand{\ee}{\end{equation}} 
\newcommand{\bea}{\begin{eqnarray}\displaystyle}
\newcommand{\eea}{\end{eqnarray}}
\newcommand{\bt}{\begin{tabular}}
\newcommand{\et}{\end{tabular}}
\newcommand{\bs}{\begin{split}}
\newcommand{\es}{\end{split}}

\newcommand{\I}{\text{I}}
\newcommand{\II}{\text{II}}

\renewcommand{\a}{\alpha}	
\renewcommand{\b}{\beta}
\newcommand{\g}{\gamma}		
\newcommand{\G}{\Gamma}
\renewcommand{\d}{\delta}
\newcommand{\D}{\Delta}
\renewcommand{\c}{\chi}			
\newcommand{\C}{\Chi}
\newcommand{\p}{\psi}			
\renewcommand{\P}{\Psi}
\newcommand{\s}{\sigma}		
\renewcommand{\S}{\Sigma}
\renewcommand{\t}{\tau}		
\newcommand{\e}{\epsilon}
\newcommand{\n}{\nu}
\newcommand{\m}{\mu}
\renewcommand{\r}{\rho}
\renewcommand{\l}{\lambda}

\newcommand{\nn}{\nonumber\\} 		
\newcommand{\newotimes}{}  				
\newcommand{\diff}{\,\text{d}}		
\newcommand{\h}{{1\over2}}				
\newcommand{\Gf}[1]{\G \Big{(} #1 \Big{)}}	
\newcommand{\floor}[1]{\left\lfloor #1 \right\rfloor}
\newcommand{\ceil}[1]{\left\lceil #1 \right\rceil}

\def\cA{{\cal A}} \def\cB{{\cal B}} \def\cC{{\cal C}}
\def\cD{{\cal D}} \def\cE{{\cal E}} \def\cF{{\cal F}}
\def\cG{{\cal G}} \def\cH{{\cal H}} \def\cI{{\cal I}}
\def\cJ{{\cal J}} \def\cK{{\cal K}} \def\cL{{\cal L}}
\def\cM{{\cal M}} \def\cN{{\cal N}} \def\cO{{\cal O}}
\def\cP{{\cal P}} \def\cQ{{\cal Q}} \def\cR{{\cal R}}
\def\cS{{\cal S}} \def\cT{{\cal T}} \def\cU{{\cal U}}
\def\cV{{\cal V}} \def\cW{{\cal W}} \def\cX{{\cal X}}
\def\cY{{\cal Y}} \def\cZ{{\cal Z}}

\def\mC{\mathbb{C}} \def\mP{\mathbb{P}}  
\def\mR{\mathbb{R}} \def\mZ{\mathbb{Z}} 
\def\mT{\mathbb{T}} \def\mN{\mathbb{N}}
\def\mH{\mathbb{H}} \def\mX{\mathbb{X}}
\def\CP{\mathbb{CP}}
\def\RP{\mathbb{RP}}
\def\Z{\mathbb{Z}}
\def\N{\mathbb{N}}
\def\H{\mathbb{H}}

\newcommand{\MyBlue}{\color [rgb]{0,0,0.8}}
\def\BG#1{{\MyBlue [#1]}}

\def\b{\bigskip}

\begin{flushright}
\end{flushright}
\vspace{20mm}
\begin{center}
{\LARGE Bootstrapping the effect of the twist operator\\\vspace{2mm} in the D1D5 CFT}
\\
\vspace{18mm}
\textbf{Bin} \textbf{Guo}{\footnote{bin.guo@ipht.fr}}~\textbf{and} ~ \textbf{Shaun}~  \textbf{Hampton}{\footnote{shaun.hampton@ipht.fr}}
\\
\vspace{10mm}

${}$Institut de Physique Th\'eorique,\\
	Universit\'e Paris-Saclay,
	CNRS, CEA, \\ 	Orme des Merisiers,\\ Gif-sur-Yvette, 91191 CEDEX, France  \\

\vspace{8mm}
\end{center}

\vspace{4mm}

\thispagestyle{empty}

\begin{abstract}

In the D1D5 CFT the twist operator of order 2 can twist together two copies in the untwisted sector into a single joined copy in the twisted sector. Traditionally, this effect is computed by using the covering map method. 
Recently, a new method was developed using the Bogoliubov ansatz and conformal symmetry to compute this effect in a toy model of one free boson. In this paper, we use this method with superconformal symmetry to compute the effect of the twist operator in the D1D5 CFT. This may provide more effective tools for computing correlation functions of twist operators in this system.

\vspace{3mm}

\end{abstract}
\newpage

\setcounter{page}{1}

\numberwithin{equation}{section} 

\tableofcontents

\newpage

\section{Introduction}

The D1D5 CFT is widely used in AdS$_3$/CFT$_2$ as a boundary theory to understand the bulk physics \cite{Maldacena:1997re,Strominger:1996sh,Maldacena:1999bp,Seiberg:1999xz,Dijkgraaf:1998gf,Larsen:1999uk,Jevicki:1998bm,deBoer:1998kjm,Martinec:2022ofs,Balthazar:2021xeh,Bufalini:2022wyp,Eberhardt:2018ouy}.
It has yielded many interesting results related to black holes \cite{Callan:1996dv,Das:1996wn,Das:1996ug,Maldacena:1996ix}. It provides a system in which to count black hole microstates \cite{Strominger:1996sh,Maldacena:1999bp} of the corresponding gravitational theory in the holographic dual. In certain cases these microstates are explicitly known \cite{Lunin:2001fv,Lunin:2001jy,Mathur:2005zp,Bena:2015bea,Bena:2016agb,Bena:2016ypk,Bena:2017xbt,Ceplak:2018pws,Heidmann:2019zws,Ganchev:2021iwy,Ganchev:2021pgs,Bena:2022sge} and correspond to particular states in the CFT \cite{Kanitscheider:2006zf,Kanitscheider:2007wq,Taylor:2007hs,Giusto:2015dfa,GarciaTormo:2019inl,Giusto:2019qig,Rawash:2021pik,Ganchev:2021ewa}. A simple description of the CFT  exists at what is known as `orbifold' point  where the theory is a free symmetric product orbifold \cite{Seiberg:1999xz,Dijkgraaf:1998gf,Larsen:1999uk,Jevicki:1998bm,deBoer:1998kjm,Pakman:2009zz,Pakman:2009ab,Pakman:2009mi,Lunin:2000yv,Lunin:2001pw,Borisov:1997nc}
$\mathcal{M}^N/S_{N}$ where $N=N_1N_5$ with $N_1$ the number of $N_1$ branes and $N_5$ the number of $N_5$ branes 
and $S_N$ is the permutation group.
However, the supergravity description lives far away from this point in the moduli space. Therefore, only quantities which are protected, as one flows from the free theory into the supergravity regime, can be reliably computed. To gain more understanding about their relationship one can add a deformation \cite{David:1999ec,Gomis:2002qi,Gava:2002xb,Avery:2010er,Avery:2010hs,Avery:2010qw,Burrington:2014yia,Eberhardt:2021vsx} which connects the two points. More recently, there has been some debate about which operator flows to the supergravity point from the study of tensionless string in AdS$_3$\cite{Eberhardt:2019ywk,Eberhardt:2020akk,Dei:2020zui,Knighton:2020kuh,Eberhardt:2018ouy,Gaberdiel:2021kkp,Eberhardt:2021vsx}. In any case, it is clear that there is one ingredient which is critical in this process. That is the twist operator of order 2. The D1D5 system is composed of component strands with a total length of $N=N_1N_5$. The twist operator can join and separate these strands. A challenge, from the CFT perspective, is using this twist operator to move far away from the free point. This is because computations typically involve introducing covering maps, which quickly become complicated as one adds more twists. For additional works involving deformations away from the orbifold point see \cite{Carson:2014yxa,Carson:2014xwa,Carson:2014ena,Carson:2016cjj,Carson:2016uwf,Carson:2015ohj,Carson:2017byr,Hampton:2019csz,Guo:2021ybz,Guo:2021gqd,Hampton:2019hya,Hampton:2018ygz,Guo:2019pzk,Guo:2019ady,Guo:2020gxm,Benjamin:2021zkn,Apolo:2022fya,Gaberdiel:2015uca,Martinec:2020cml,Ceplak:2021kgl,Lima:2020boh,Lima:2020nnx,Lima:2020urq,Lima:2020kek,Lima:2021wrz,AlvesLima:2022elo,Guo:2022ifr}.  

Alternative methods to using just the covering maps to compute the effects of the twist operator, is to bootstrap the computations by using Bogoliubov transformations and conformal symmetry. There have been several works along these lines. In \cite{Carson:2014xwa} the authors used a Bogoliubov transformation to compute various coefficients coming from the effects of the twist operator. This method involved inverting infinite dimensional matrices. The authors in \cite{Burrington:2014yia} used a combination of conformal symmetry and covering maps to compute various effects produced by the twist operator. Recently, in \cite{Guo:2022sos}, a method was developed for orbifold CFTs of a single boson, which used a weak Bogoliubov ansatz and conformal symmetry to determine the effect of the twist operator completely. This paper is a continuation of this method to include fermions and thereby compute the full effect of the twist operator in the D1D5 CFT purely by using the symmetries of the theory. It is hopeful that these bootstrap techniques can be applied to higher orders in the twist operator which in turn can help to describe effects and processes which are far from the orbifold point. This would provide insight into nonperturbative processes in the D1D5 CFT which would correspond to processes in the supergravity regime. 

This paper is organized as follows. In section \ref{sec D1D5} we review the D1D5 CFT. In section \ref{sec effect} we describe the effect of the twist operator and introduce the `weak' Bogoliubov ansatz. In section \ref{sec fermion} we compute the effect for fermionic modes. In section \ref{susy} we compute the effect for bosonic modes by using supersymmetry. In section \ref{sec solution} we summarize all the coefficients we obtained for the effect of the twist operator. In section \ref{discussion} we discuss our results and outlook.

\section{The D1D5 CFT}\label{sec D1D5}

In this section we review the D1D5 CFT at the orbifold point. Consider type IIB string theory compactified on $M_{4,1}\times S^1\times T^4$ with $N_1$ D1 branes wrapping $S^1$ and $N_5$ D5 branes wrapping $S^1\times T^4$. The torus is taken to be much smaller than $S^1$ and thus the low energy limit of this configuration reduces to a 1+1 $\mathcal{N}=4$ Super Conformal Field Theory (SCFT) living on a circle of radius $R$, with central charge $c=6N$ where $N=N_1N_5$. The theory has $SO_E(4)\cong SU(2)_L\times SU(2)_R$ symmetry in the noncompact directions where E is stands for 'external'. The torus has an $SO_I(4)\cong SU_1(2)\times SU_2(2)$ symmetry where I stands for `internal'. Upon compactification, this symmetry is broken however it serves as a useful organizing principle. The base space is a cylinder parameterized by the rescaled coordinates.
\bea
\t\!&=&\! it/R,\quad -\infty<\t<\infty\cr
\s\!&=&\! y/R,\quad 0\leq \sigma \leq 2\pi
\eea
Here $t$ is the physical Lorenztian time and $y$ is the physical coordinate of $S^1$. We can group these coordinates into a single complex coordinate
\bea
w = \t + i\sigma
\eea
The theory contains four bosons and four fermions.
The bosonic fields can be grouped together according to the $SU_1(2)\times SU_2(2)$ symmetry as $X_{A\dot A}$ where $A,\dot A = +,-$. The left moving fermions can be grouped together according to the combined symmetries $SU_L(2)\times SU_1(2)$ as $\psi^{\a A}$ where $\a = +,-$. The right moving fields are grouped in a similar manner. 

This CFT is a symmetric product orbifold $\mathcal{M}^N/S_N$, of $N$ copies of a `seed' CFT $\mathcal{M}$ where $\mathcal{M}$ is the target space and $S_N$ is the permutation group. This CFT contains various twisted sectors which are obtained by applying a twist operator which joins and splits copies of the CFT by altering their boundary conditions. To better illustrate this idea consider $N$ bosons $X^{(j)}_{A\dot A}$ where $j=1,2,\ldots,N$. Consider a twist operator of order $k$ 
which will twist together $k$ out of $N$ copies, e.g. the first $k$ copies. 
As you rotate by an angle $2\pi$ around the cylinder the action of the twist on the bosons gives
\bea 
X^{(1)}_{A\dot A}\to X^{(2)}_{A\dot A}\to X^{(3)}_{A\dot A}\to\ldots\to X^{(k)}_{A\dot A}\to X^{(1)}_{A\dot A}
\eea
for copies $1$ to $k$. For the remaining copies we have
\bea
X^{(j)}_{A\dot A}\to X^{(j)}_{A\dot A},\quad k+1\leq j\leq N
\eea
In this paper, we take the fermions to be in the Ramond sector.
Under the rotation, we have
\bea 
\psi^{\a A (1)}\to \psi^{\a A (2)}\to \psi^{\a A (3)}\to\ldots\to \psi^{\a A (k)}\to \psi^{\a A (1)}
\eea
for copies $1$ to $k$ and for the remaining copies we have
\bea
\psi^{\a A (j)}\to \psi^{\a A(j)},\quad k+1\leq j\leq N
\eea 
Thus it is convenient to define a single field $X_{A\dot A}$ on the $k$-wound copy. The field $X_{A\dot A}$ equals to $X^{(j)}_{A\dot A}$ on the $j$-th segment of the $k$-wound copy, i.e.
\be
X_{A\dot A}(\tau, \s ) = X^{(j)}_{A\dot A}(\tau,\s ), ~~~2\pi(j-1) \leq \s < 2\pi j
\ee
The field $X_{A\dot A}$ has the boundary condition
\be
X_{A\dot A}(w + 2\pi k i) = X_{A\dot A}(w)
\ee
Similarly we define a single field $\psi^{\alpha A}$ on the $k$-wound copy as follows
\be
\psi^{\alpha A}(\tau, \s ) = \psi^{\alpha A(j)}(\tau,\s ), ~~~2\pi(j-1) \leq \s < 2\pi j
\ee
with the boundary condition
\be
\psi^{\alpha A}(w + 2\pi k i) = \psi^{\alpha A}(w)
\ee
Because of these boundary conditions, one can define modes on the $k$-wound copy on the cylinder for bosons on a constant time slice as
\bea
\a_{A\dot A,{n\over k }} = {1\over2\pi }\int_{\s=0}^{2\pi k}dw\, e^{{n\over k}w}\partial X_{A\dot A}(w)
\eea
and for fermions in the R sector as 
\bea
d^{\a A}_{n\over k} = {1\over2\pi i}\int_{\s=0}^{2\pi k}dw\, e^{{n\over k}w} \psi^{\a A}(w)
\eea
where $n$ is an integer.
In the next section we discuss the effect of the twist operator in the D1D5 CFT.

\section{The effect of the twist operator}\label{sec effect}

Here we outline the effect of the twist operator $\s^+_2$ in the D1D5 CFT.  
The operator $\s^+_2$ has twist order 2 and is a chiral primary with dimension and charge $h=j=1/2$. 
In the rest of the paper we will omit the subscript $2$ and label it as $\s^+$ for brevity.
The twist operator changes the vacuum of the CFT. Anytime such a change happens, one can introduce a Bogoliubov transformation which relates modes defined with respect to the original vacuum to modes defined with respect to the modified vacuum. For more work on the twist operator in the D1D5 CFT see \cite{Burrington:2017jhh,Burrington:2018upk,Arutyunov:1997gt,Arutyunov:1997gi,deBoer:1998kjm,Dei:2019iym}.

In this paper we take $N=2$ where the initial vacuum is composed of two singly wound copies of the D1D5 CFT in the Ramond sector on the cylinder. Consider an initial state, $|\phi\rangle$ which contains an arbitrary number of bosonic and fermionic modes acting on this vacuum
\bea
|\phi\rangle=\a^{(i_1)}_{A_1\dot A_1,-m_1}\a^{(i_2)}_{A_2\dot A_2,-m_2}\ldots \a^{(i_k)}_{A_k\dot A_k,-m_k}d^{\beta_1B_1(j_1)}_{-n_1}d^{\beta_2B_2(j_2)}_{-n_2}\ldots d^{\beta_lB_l(j_l)}_{-n_l}|0^-_R\rangle^{(1)}|0^-_R\rangle^{(2)}
\eea
where $i_{k'},j_{l'}=1,2$ are copy labels and $A_{k'},\dot A_{k'},B_{l'},\dot B_{l'}=+,-$ and $\beta_{l'}=+,-$. Each copy of the singly wound vacuum carries the quantum numbers $h=1/4$ and $j=-1/2$. We introduce the twist, $\s^+(w)$ at some point $w$ on the cylinder. Acting it on the state $|\phi\rangle$ will join the two singly wound vacua into a doubly wound vacuum and produce three basic effects.

\bigskip

(i) Contraction: The action of the twist can cause any two bosonic modes, 
\bea \a^{(i)}_{A\dot A,-m} \text{ and } \a^{(j)}_{B\dot B,-n}
\eea
and any two fermionic modes
\bea 
d^{\a A(i)}_{-m}\text{ and } d^{\beta B(j)}_{-n}
\eea 
with the appropriate charges to `Wick' contract producing a number
\bea
\text{bosons}&:& C^{ij}_{A\dot A C\dot C}[n_1,n_2]\equiv C[\a^{(i)}_{A\dot A,-n_1}\a^{(j)}_{C\dot C,-n_2}]\equiv \e_{AC}\e_{\dot A\dot C}C^{ij}_B[n_1,n_2]\cr 
\cr
\text{fermions}&:& C_F^{ij,\a A\beta B}[n_1,n_2]\equiv C[d^{\a A(i)}_{-n_1} d^{\beta B(j)}_{-n_2}]\equiv \e^{\alpha\beta}\e^{AB}C^{ij}_F[n_1,n_2]
\eea 
For the contraction we consider all possible pairs of bosons and all possible pairs of fermions. They can contract together as described above but if not they can also pass through the twist as we will describe below.

\bigskip 

(ii)Propagation: Any mode remaining after contraction produces a linear combination of modes on a doubly wound string weighted by a certain set of coefficients
\bea\label{p rule}
\a^{(i)}_{A\dot A,-n}&\to&\sum_{p> 0}f_i[-n,-p]\a_{A\dot A,-p},~~~~n>0\cr
d^{+ A(i)}_{-n}&\to&\sum_{p\geq 0}f^{+}_i[-n,-p]d^{+ A}_{-p},~~~~n\geq 0\cr
d^{- A(i)}_{-n}&\to&\sum_{p> 0}f^{-}_i[-n,-p]d^{- A}_{-p},~~~~n>0
\eea
where $n$ is an integer. In appendix \ref{app global}, we show that
\bea \label{f hi}
f_i[-n,-n] &=& f^{\pm}_i[-n,-n] = {1\over2}
\cr
\cr
f_i[-n,-p],~f^{\pm}_i[-n,-p]&\neq& 0 \quad\text{ for } p\neq n \text{ and } p \text{ a positive half integer }
\eea 

\bigskip

(iii) Pair creation: After the previous two steps, the twist acts on two copies of the singly wound vacuum. The effect is given below
\bea\label{chi}
|\chi\rangle&\equiv& \sigma^+(w)|0^-_R\rangle^{(1)}|0^-_R\rangle^{(1)}\cr
&=&\exp\big(\sum_{m,n>0}\gamma^B_{mn}[-\a_{++,-m}\a_{--,-n} + \a_{+-,-m}\a_{-+,-n}]\big)\cr
&&\exp\big( \sum_{m,n>0}\gamma^F_{mn}[d^{++}_{-m}d^{--}_{-n} - d^{+-}_{-m}d^{-+}_{-n}] \big)|0^{2-}_R\rangle
\eea
Where $\gamma^B_{mn},\gamma^F_{mn}$ are the Bogoliubov coefficients describing pair creation for bosons and fermions respectively. In appendix \ref{app global}, we show that 
\bea\label{g hi}
\gamma^B_{mn},\gamma^F_{mn}&\neq& 0,~~~~~\quad  m,n>0 ~~ \text{ and }~~ m,n \text{ half integer }
\eea

Since the coefficients in the contraction, propagation and pair creation are independent of each other, we call it `weak' the Bogoliubov ansatz as in \cite{Guo:2022sos}.
To better understand these rules, we give two examples in the following.
Applying the twist operator to a single fermionic mode in the initial state gives
\bea\label{single mode}
\s^+(w)d^{- A(i)}_{-n}|0^-_R\rangle^{(1)}|0^-_R\rangle^{(2)} = \sum_{p>0}f^{-}_i[-n,-p]d^{- A}_{-p}|\chi\rangle
\eea
where the initial mode passes through the twist operator using the rule (\ref{p rule}). Then the twist operator acts on the untwisted vacuum to produce pairs of modes using the rule (\ref{chi}).
Applying the twist operator to two fermionic modes in the initial state, we obtain
\bea\label{two modes}
&&\s^+(w)d^{+ A(i)}_{-n_1}d^{- B(j)}_{-n_2}|0^-_R\rangle^{(1)}|0^-_R\rangle^{(2)}\cr 
&&= \bigg(\sum_{p\geq0}f^{+}_i[-n_1,-p]d^{+ A}_{-p}\sum_{q>0}f^{-}_j[-n_2,-q]d^{- B}_{-q} + \e^{+-}\e^{AB}C^{ij}_F[n_1,n_2]\bigg)|\chi\rangle
\eea
The first term comes from the propagation of the two initial modes while the second term comes from the contraction. The state $|\chi\rangle$ comes from the pair creation.

\section{Bootstrapping the effect of the twist operator}\label{sec fermion}

In this section we will take the weak Bogoliubov ansatz and apply the superconformal generators to get recursion relations for the coefficients in the ansatz. We will start with the fermions and compute for the bosons in the next section using supersymmetry.

\subsection{Pair creation}

In this subsection, we will derive the pair creation coefficients, $\gamma^F_{mn}$. 
Starting with the state
\be
\left(L_{-1}+J^3_{-1}\right)|0^-_R\rangle^{(1)}|0^-_R\rangle^{(2)}=0
\ee
we then apply the twist operator $\sigma^+$ to find some recursion relations for $\gamma^F_{mn}$. 
Consider the state
\bea\label{gammaF}
0&=&\sigma^+(w)(L_{-1}+ J^3_{-1})|0^-_R\rangle^{(1)}|0^-_R\rangle^{(2)}\nn
&=&\left((L_{-1}+J^3_{-1})\sigma^+(w)-[L_{-1}+J^3_{-1},\sigma^+(w)]\right)|0^-_R\rangle^{(1)}|0^-_R\rangle^{(2)}
\eea
In the second term, the commutator is given by
\bea\label{Lm1}
[L_{-1},\sigma^+(w)] &=& \oint_{C^{(4\pi)}_{w}} \frac{dw'}{2\pi i}e^{-w'} T(w')  \sigma^+(w)\cr
&=&e^{-w}\oint_{C^{(4\pi)}_{w}} \frac{dw'}{2\pi i}e^{-(w'-w)} T(w')  \sigma^+(w)\cr
&=&e^{-w}\oint_{C^{(4\pi)}_{w}} \frac{dw'}{2\pi i}(1-(w'-w)+\frac{1}{2}(w'-w)^2+\dots) T(w')
\sigma^+(w)\cr
&=&e^{-w}(L_{-1}^{(w)}-L_0^{(w)}+\frac{1}{2}L_1^{(w)}+\dots)  \sigma^+(w)\cr
&=&e^{-w}\bigg(\partial-\frac{1}{2}\bigg)\sigma^+(w)
\eea
and
\bea\label{Jm1}
[J^3_{-1},\sigma^+(w)] &=&\oint_{C^{(4\pi)}_{w}} \frac{dw'}{2\pi i}e^{-w'} J^3(w') \sigma^+(w)\cr
&=&e^{-w} \oint_{C^{(4\pi)}_{w}} \frac{dw'}{2\pi i}e^{-(w'-w)} J^3(w') \sigma^+(w)\cr
&=&e^{-w}\oint_{C^{(4\pi)}_{w}} \frac{dw'}{2\pi i}(1-(w'-w)+\ldots)J^3(w')\sigma^+(w)\cr
&=&e^{-w}( J^{3,(w)}_0 -  J^{3,(w)}_{1} + \ldots ) \sigma^+(w)\cr
&=&e^{-w}\frac{1}{2}\sigma^+(w)
\eea
where we have used 
\bea\label{LJsigma}
&&L^{(w)}_{-1}\sigma^+(w) = \partial\sigma^+(w),~~L^{(w)}_{0}\sigma^+(w) = \frac{1}{2}\sigma^+(w),~~L^{(w)}_{n>0}\sigma^+(w) = 0 \nn
&&J^{(w)}_{0}\sigma^+(w) = \frac{1}{2}\sigma^+(w),~~J^{(w)}_{n>0}\sigma^+(w)  = 0
\eea
Inserting the relations (\ref{Lm1}) and (\ref{Jm1}) into (\ref{gammaF}) and using the pair creation ansatz (\ref{chi}) gives 
\bea\label{eq gamma}
0&=&\sigma^+(w)(L_{-1}+ J^3_{-1})|0^-_R\rangle^{(1)}|0^-_R\rangle^{(2)}\nn
&=&\left(L_{-1}+J^3_{-1}-e^{-w}\partial\right) \text{exp}\bigg[\sum_{m,n\geq 0}\gamma^F_{m+1/2,n+1/2} d^{++}_{-(m+1/2)}d^{--}_{-(n+1/2)} \bigg] |0^{2-}_R\rangle
\eea
where $m$ and $n$ are non-negative integers. Note that in the exponential we only keep the fermionic modes in the $++,--$ sector because here they are the only modes which are necessary to derive the coefficient $\gamma^F_{m+1/2,n+1/2}$.   
We will show that this equation determines the $\gamma^F_{m+1/2,n+1/2}$ completely.
Let us look for the following term in (\ref{eq gamma})
\be
d^{++}_{-(m+1/2)}d^{--}_{-(n+1/2)}|0^{2-}_R\rangle
\ee

\subsubsection{The recursion relations}

For $m,n>0$, we obtain the relation
\bea\label{gamma dd}
&&\gamma^F_{m-1/2,n+1/2}\left[L_{-1}+J^3_{-1},d^{++}_{-(m-1/2)}\right]d^{--}_{-(n+1/2)}\nn
&+&\gamma^F_{m+1/2,n-1/2}d^{++}_{-(m+1/2)}\left[L_{-1}+J^3_{-1},d^{--}_{-(n-1/2)}\right]\nn
&-&e^{-w}\partial \gamma^F_{m+1/2,n+1/2} d^{++}_{-(m+1/2)}d^{--}_{-(n+1/2)}=0
\eea
Using the commutators
\bea
\left[L_{-1}+J^3_{-1},d^{++}_{-(m-1/2)}\right]=\left(m+1/2\right)d^{++}_{-(k+3/2)}\nn
\left[L_{-1}+J^3_{-1},d^{--}_{-(n-1/2)}\right]=\left(n-1/2\right)d^{--}_{-(k+3/2)}
\eea
we obtain a recursion relation
\be\label{gamma r1}
\gamma^F_{m-1/2,n+1/2}(m+1/2)
+\gamma^F_{m+1/2,n-\frac{1}{2}}(n-1/2)
=e^{-w}\partial \gamma^F_{m+1/2,n+1/2} ~~~~~~~m,n>0
\ee

For the case $m=0,n>0$, we do not have the first term in (\ref{gamma dd}).
Thus we have
\be\label{gamma r2}
\gamma^F_{1/2,n-1/2}(n-1/2)
=e^{-w}\partial \gamma^F_{1/2,n+1/2} ~~~~~~~m=0,n>0
\ee

For the case $m>0,n=0$, we do not have the second term in (\ref{gamma dd}).
Thus we have
\be\label{gamma r3}
\gamma^F_{m-1/2,1/2}(m+1/2)
=e^{-w}\partial \gamma^F_{m+1/2,1/2} ~~~~~~~m>0,n=0
\ee

For the case $m=n=0$, we have
\be
\left(L_{-1}+J^3_{-1}\right)|0^{2-}_R\rangle-e^{-w}\partial \gamma^F_{1/2,1/2} d^{++}_{-1/2}d^{--}_{-1/2}|0^{2-}_R\rangle=0
\ee
Using
\be\label{mn0}
\left(L_{-1}+J^3_{-1}\right)|0^{2-}_R\rangle=-\frac{1}{4}d^{++}_{-1/2}d^{--}_{-1/2}|0^{2-}_R\rangle
\ee
we obtain
\be\label{eq gamma hh}
 \partial \gamma^F_{1/2,1/2} = -\frac{e^{w}}{4}
\ee

\subsubsection{The solution}
To determine the initial condition for this system of differential equations, we consider
\bea
|0^{2-}_R\rangle = \s^+(w\to -\infty)|0^{-}_R\rangle^{(1)}|0^{-}_R\rangle^{(2)}
\eea
since the twist operator $\s^+$ is the lowest dimension operator that changes the untwisted sector to the twisted sector. 
Thus from the ansatz (\ref{chi}) we have
\be
\gamma^F_{m+1/2,n+1/2}(w\to -\infty) = 0
\ee
The different equations can be solved using these initial conditions. The solution to (\ref{eq gamma hh}) is
\be\label{gammahh}
\gamma^F_{1/2,1/2} = -\frac{e^w}{4}
\ee
We can find all other $\gamma_{m+1/2,n+1/2}$'s by using the relations (\ref{gamma r1}), (\ref{gamma r2}), and (\ref{gamma r3}). The $\gamma_{m+1/2,n+1/2}$'s form an inverted triangle, where the space between each lattice point is an integer and the bottom lattice point is $\gamma_{1/2,1/2}$. The relation (\ref{gamma r2}) moves you along the left edge, (\ref{gamma r3}) moves you along the right edge and (\ref{gamma r1}) moves you within the interior as follows.
\be
\begin{tikzcd}[row sep=16pt, column sep=-5pt]
~~~~~~& &~~~~~~~~ & &~~~~~~~~ & & ~~~~~~ \\
 &\arrow[ul,dotted]\gamma^F_{1/2,5/2}\arrow[ur,dotted]& &\arrow[ul,dotted]\gamma^F_{3/2,3/2} \arrow[ur,dotted] & &\arrow[ul,dotted]\gamma^F_{5/2,1/2}\arrow[ur,dotted] & \\
 & &\arrow[ul]\gamma^F_{1/2,3/2}\arrow[ur] &  &\arrow[ul]\gamma^F_{3/2,1/2}\arrow[ur] & &\\
 & & & \arrow[ul]\gamma^F_{1/2,1/2} \arrow[ur]& & &
\end{tikzcd}
\ee
The solution is
\bea\label{gamma} 
\gamma^F_{m+{1\over2},n+{1\over2}} &=& -{e^{(m+n+1)w}\Gamma[{3\over2}+m]\Gamma[{3\over2}+n]\over(2n+1)\pi(m+n+1)\Gamma[m+1]\Gamma[n+1]}
\eea
where $m$ and $n$ are non-negative integers. 

\subsection{Propagation}\label{propagation}

In this subsection, we derive the expressions for propagation $f^{\pm}_i [-n,-p]$ which correspond to a fermionic mode passing through the twist operator.

\subsubsection{Relations using $L_0 + J^3_0$}

Here we use the generator $L_0 + J^3_0$ to find the $w$-dependence of $f^-_i[-1,-p]$. We start with
\bea
 (L_0 + J^3_0)d^{--(i)}_{-1} |0^-_R\rangle^{(1)}|0^-_R\rangle^{(2)} = 0
\eea
Then applying the twist operator, we obtain the relation
\bea
0&=& \s^+_2(w) (L_0 + J^3_0)d^{--(i)}_{-1} |0^-_R\rangle^{(1)}|0^-_R\rangle^{(2)}
\eea
Now commuting $L_0 + J_0^3$ to the left gives 
\bea\label{L0}
0&=& \left((L_0 + J^3_0) \s^+_2(w)  - [L_0 + J^3_0,\s^+_2(w) ]\right) d^{--(i)}_{-1} |0^-_R\rangle^{(1)}|0^-_R\rangle^{(2)}
\eea
The commutator is given by
\bea 
[L_{0}+J^3_{0},\sigma^+(w)] &=& \bigg(\oint_{C^{(4\pi)}_{w}} \frac{dw'}{2\pi i} T(w') + \oint_{C^{(4\pi)}_{w}} \frac{dw'}{2\pi i} J^3(w')\bigg) \sigma^+(w)\cr
&=&(L^{(w)}_{-1} + J^{3,(w)}_0) \sigma^+(w)\cr
&=&\bigg(\partial + {1\over2}\bigg)\sigma^+(w)
\eea
Inserting this into (\ref{L0}) gives
\bea
0=\bigg(L_0 + J^3_0-\bigg(\partial + {1\over2}\bigg)\bigg) \s^+_2(w) d^{--(i)}_{-1} |0^-_R\rangle^{(1)}|0^-_R\rangle^{(2)}
\eea
Using the ansatz (\ref{single mode}) and keeping terms with only one $d^{--}$ mode we get
\bea
0&=&\bigg(L_0 + J^3_0-\bigg(\partial + {1\over2}\bigg)\bigg)\sum_{p>0} f_i^-[-1,-p]d^{--}_{-p} |0^{2-}_R\rangle
\eea
Acting with $L_0 + J^3_0$, the partial derivative and matching coefficients we obtain 
\bea
(p-1)f_i^-[-1,-p] =\partial f_i^-[-1,-p]  
\eea
This relation implies that $f_i^-[-1,-p]$ will take the following functional form
\bea\label{w_dep}
f_i^-[-1,-p]\propto e^{(p-1)w}
\eea
Using this result in the following section we compute the exact form of $f_i^-[-1,-p]$ which includes the proportionality constant.  

\subsubsection{Relations using $J^3_1$}

Here we use $J^3_1$ to compute the full expression for $f_i^-[-1,-p]$ upto an over constant $C$. We begin with the
state
\bea
J^3_1d^{--(i)}_{-1}|0^-_R\rangle^{(1)}|0^-_R\rangle^{(2)}=0
\eea 
Applying the twist operator gives
\bea\label{start f-}
0=\s^+(w)J^3_1d^{--(i)}_{-1}|0^-_R\rangle^{(1)}|0^-_R\rangle^{(2)}
\eea 
Commuting $J^3_1$ through the twist gives the relation
\bea \label{L1}
0=(J^3_1 \s^+(w) - [J^3_1 , \s^+(w)])d^{--(i)}_{-1}|0^-_R\rangle^{(1)}|0^-_R\rangle^{(2)} 
\eea
where the commutator is given by
\bea \label{cj3}
[J^3_{1},\sigma^+(w)] &=& \oint_{C^{(4\pi)}_{w}} \frac{dw'}{2\pi i} e^{w'}J^3(w') \sigma^+(w)\cr
&=& e^w \oint_{C_{w}} \frac{dw'}{2\pi i} e^{w'-w}J^3(w') \sigma^+(w)
\cr
&=& e^w \oint_{C_{w}} \frac{dw'}{2\pi i}(1 + (w'-w) + \dots)J^3(w')\sigma^+(w)
\cr
&=&e^w (J^{3,(w)}_0+J^{3,(w)}_1+\dots) \sigma^+(w)\cr
&=&{1\over2}e^w\sigma^+(w)
\eea
where we have used (\ref{LJsigma}).
Inserting this into (\ref{L1})
\bea
0=(J^3_1-{1\over2}e^w)\s^+(w)d^{--(i)}_{-1}|0^-_R\rangle^{(1)}|0^-_R\rangle^{(2)}
\eea
Again using the ansatz (\ref{chi}) and keeping only the fermionic modes we find
\bea
0&=&(J^3_1 -{1\over2}e^w )\sum_{p>0} f_i^-[-1,-p]d^{--}_{-p}\cr
&&\text{exp}\bigg[\sum_{m,n\geq 0}\gamma^F_{m+1/2,n+1/2}(d^{++}_{-(m+1/2)}d^{--}_{-(n+1/2)} - d^{+-}_{-(m+1/2)}d^{-+}_{-(n+1/2)} ) \bigg] |0^{2-}_R\rangle
\eea
Looking for the term $d^{--}_{-(n+1/2)}|0^{2-}_R\rangle$, we obtain
\bea
0&=& f_i^-[-1,-1/2]\gamma^F_{1/2,n+1/2}\big\{[J_1^3,d^{--}_{-1/2}],d^{++}_{-1/2}\big\}d^{--}_{-(n+1/2)} \cr
&&+ f_i^-[-1,-(n+1/2)]\gamma^F_{1/2,1/2}d^{--}_{-(n+1/2)}\big\{[ J^3_1,d^{++}_{-1/2}],d^{--}_{-1/2}\big\}\cr
&&-f_i^-[-1,-(n+1/2)]\gamma^F_{1/2,1/2}d^{--}_{-(n+1/2)}\big\{[ J^3_1,d^{+-}_{-1/2}],d^{-+}_{-1/2}\big\}\cr
&&+f^-_i[-1,-(n+3/2)][J^3_1,d^{--}_{-(n+3/2)}]-{1\over2}e^wf^-_i[-1,-(n+1/2)]d^{--}_{-(n+1/2)}
\eea
which gives
\bea
0&=&f_i^-[-1,-1/2]\gamma^F_{1/2,n+1/2}-2f_i^-[-1,-(n+1/2)]\gamma^F_{1/2,1/2}\nn
&&-\frac{1}{2}f^-_i[-1,-(n+3/2)]-{1\over2}e^w f^-_i[-1,-(n+1/2)]
\eea
Notice that the second and fourth terms on the RHS cancel each other because of the value of $\gamma^F_{1/2,1/2}$ (\ref{gammahh}). Plugging in $\gamma^F_{1/2,n + 1/2}$ from (\ref{gamma}), we obtain
\be\label{fmi}
f_i^-[-1,-(n+3/2)] =  -{2e^{(n+1)w}\Gamma[{3\over2}]\Gamma[{3\over2}+n]\over(2n+1)\pi\Gamma[n+2]}f_i^-[-1,-1/2],~~~~n\geq 0
\ee
We note that copy 2 quantities are related to copy 1 quantities by a shift of $w\to w+2\pi i$. Therefore using (\ref{w_dep}) for $f_i^-[-1,-1/2]$ we get the expression
\bea \label{f- h}
f_i^-[-1,-1/2] = C(-1)^{i+1}e^{-w/2}
\eea
For higher modes we find that 
\bea \label{f-1}
f_i^-[-1,-(n+1/2)] =  C(-1)^{i}{2e^{(n-1/2)w}\Gamma[{3\over2}]\Gamma[{1\over2}+n]\over\pi(2n-1)\Gamma[n+1]},~~~~n\geq 0
\eea
In appendix \ref{app f-} we compute the full expression for $f^-_i[-n,-(m+1/2)]$ which is given by 
\bea\label{full f-}
\!\!f_i^-[-n,-(m+1/2)]=C(-1)^{i}{2\Gamma[{1\over2} + n] \Gamma[{1\over2} + m]\over\pi\Gamma[n] \Gamma[1 + m]}{e^{(m - n + 1/2) w}\over2 m - 2 n + 1},~~~n>0,m\geq 0
\eea
where the constant $C$ is computed in (\ref{Const}).

\subsubsection{Relations using $J^+_1$}

In the previous section we derived the propagation $f_i^-[-n,-p]$. We can derive $f_i^+[-n,-p]$ in a similar way by replacing $d^{--}$ in (\ref{start f-}) by $d^{++}$. However, this will introduce another undetermined constant similar to the constant $C$ in (\ref{full f-}). In this section, we will derive $f_i^+[-n,-p]$ by relating it to $f_i^-[-n,-p]$ using the mode $J^+_1$. In this way, no extra undetermined constant will be introduced.
We begin with the relation
\bea\label{j+}
\s^+(w)d^{++(i)}_{-n}|0^-_R\rangle^{(1)}|0^-_R\rangle^{(2)}&=&\s^+(w)J^+_1d^{-+(i)}_{-n-1}|0^-_R\rangle^{(1)}|0^-_R\rangle^{(2)}\nn
&=&J^+_1\s^+(w)d^{-+(i)}_{-(n+1)}|0^-_R\rangle^{(1)}|0^-_R\rangle^{(2)}
\eea 
where we have used the commutator similar to the commutator (\ref{cj3})
\be
[J^+_{1},\sigma^+(w)] = e^w (J^{+,(w)}_0+J^{+,(w)}_1+\dots) \sigma^+(w) =0 
\ee
Looking for the term $d^{++}_{-(m+1/2)}$ for $m\geq 0$ from both sides of (\ref{j+}), we have
\bea
&&f^+_i[-n,-(m+1/2)]d^{++}_{-(m+1/2)} \cr
&=& f^-_i[-(n+1),-(m+3/2)] [J^+_1,d^{-+}_{-(m+3/2)}]\cr
&& - f^-_i[-(n+1),-1/2]\gamma^F_{m+1/2,1/2}\lbrace [J^+_1,d^{-+}_{-1/2}],d^{--}_{-1/2} \rbrace d^{++}_{-(m+1/2)}
\eea
which gives
\be\label{fp_from_fm}
f^+_i[-n,-(m+1/2)] = f^-_i[-(n+1),-(m+3/2)] +2 f^-_i[-(n+1),-1/2]\gamma^F_{m+1/2,1/2}
\ee
Plugging in $f^-_i$, (\ref{full f-}), and $\gamma^F_{m+1/2,1/2}$, (\ref{gamma}), we find
\bea\label{full f+}
f^+_i[-n,-(m+1/2)] = C(-1)^{i}\frac{2\Gamma[\frac{1}{2}+n]}{\pi\Gamma[1+n]}
\frac{\Gamma[\frac{3}{2}+m]}{\Gamma[1+m]}\frac{e^{(m-n+1/2)w}}{2m-2n+1}
\eea

\subsection{Contraction}\label{C}

In this section we derive the expression for the contraction $C^{ij}_F[n_1,n_2]$ in terms of the propagation $f^{\pm}_i[-n,-p]$. Let's start with the following state
\bea
|\psi\rangle\equiv\s^+_2(w)d^{++(i)}_{-n}d^{--(j)}_{-1}|0^-_R\rangle^{(1)}|0^-_R\rangle^{(2)}
\eea
where $n\geq0$.

Using our ansatz we can write $|\psi\rangle$ as 
\bea
|\psi\rangle=\big( \sum_{p> 0}f^+_i[-n,-p]d^{++}_{-p}\sum_{p'>0}f^-_j[-1,-p']d^{--}_{-p'}+C^{ij}_F[n,1]\big)|\chi\rangle
\eea
Expanding $|\psi\rangle$ and keeping only terms which contain no fermionic modes we obtain
\bea\label{psi1}
|\psi\rangle = C^{ij}_F[n,1]|0^{2-}_R\rangle  + \ldots 
\eea
Let's again consider the state $|\psi\rangle$
\bea
|\psi\rangle=\s^+_2(w)d^{++(i)}_{-n}d^{--(j)}_{-1}|0^-_R\rangle^{(1)}|0^-_R\rangle^{(2)}
\eea
We can rewrite this as
\bea 
|\psi\rangle\!\!&=&\!\!2\s^+_2(w)J^3_1d^{++(i)}_{-(n+1)}d^{--(j)}_{-1}|0^-_R\rangle^{(1)}|0^-_R\rangle^{(2)}
\cr
\!\!&=&\!\!2(J^3_1\s^+_2(w) - [J^3_1,\s^+_2(w)] )d^{++(i)}_{-(n+1)}d^{--(j)}_{-1}|0^-_R\rangle^{(1)}|0^-_R\rangle^{(2)}\cr
\cr
\!\!&=&\!\!2\bigg(J^3_1-{1\over2}e^w\bigg)\s^+_2(w)d^{++(i)}_{-(n+1)}d^{--(j)}_{-1}|0^-_R\rangle^{(1)}|0^-_R\rangle^{(2)} \cr
\!\!&=&\!\!2 \bigg(J^3_1-{1\over2}e^w\bigg) \bigg(\sum_{p> 0}f^+_i[-(n+1),-p]d^{++}_{-p}\sum_{p'>0}f^-_j[-1,-p']d^{--}_{-p'} + C_F^{ij}[n+1,1]\bigg)|\chi\rangle\nn
\eea
where in the third equality we have used (\ref{cj3}).
Expanding the above and again keeping only terms which contain no modes we obtain
\bea\label{psi2} 
|\psi\rangle\!\!&=&\!\!\big(2f^+_i[-(n+1),-1/2]f^-_j[-1,-1/2]\lbrace[J^3_1,d^{++}_{-1/2}] , d^{--}_{-1/2} \rbrace 
\cr 
&&+2C^{ij}_F[n+1,1]\gamma_{1/2,1/2}(\lbrace[J^3_1,d^{++}_{-1/2}] , d^{--}_{-1/2} \rbrace - \lbrace[J^3_1,d^{+-}_{-1/2}] , d^{-+}_{-1/2} \rbrace)
\cr
&&- e^wC^{ij}_F[n+1,1]\big)|0^{2-}_R\rangle + \ldots
\cr 
\!\!&=&\!\! \big(-2f^+_i[-(n+1),-1/2]f^-_j[-1,-1/2] -  (e^w+4\gamma_{1/2,1/2})C^{ij}_F[n+1,1] \big)|0^{2-}_R\rangle \cr
&&+ \ldots
\eea 
Notice that the second term is zero because of the value of $\gamma^F_{1/2,1/2}$ (\ref{gammahh}).
Comparing (\ref{psi1}) and (\ref{psi2}) we obtain the relation
\bea\label{C0n}
C^{ij}_F[n,1]=-2f^+_i[-(n+1),-1/2]f^-_j[-1,-1/2]
\eea
For the above expression we insert $f^+_i[-(n+1),-1/2]$ which is computed in (\ref{full f+}) and the expression for $f^-_j[-1,-1/2]$, (\ref{full f-}).   
Therefore (\ref{C0n}) becomes
\bea\label{Cn1}
C^{ij}_F[n,1] = - C^2(-1)^{i+j}e^{-(n+1)w}\frac{\Gamma[\frac{1}{2}+n]}{\sqrt\pi(1+n)\Gamma[1+n]}
\eea
In appendix \ref{app C} we derive the full expression for $C^{ij}[n_1,n_2]$ and it is given by
\bea\label{Cfull}
C^{ij}_F[n_1,n_2] =-C^2(-1)^{i+j}{2\Gamma[{1\over2}+n_1]\Gamma[{1\over2}+n_2]\over \pi n_1\Gamma[n_1]\Gamma[n_2]}{e^{-(n_1+n_2)w}\over n_1+n_2}
\eea

\subsection{Relations using $J^3_{-1}$}

We have derived all the functions in the effect of a twist operator with an undetermined constant $C$. Here we will compute this constant by using a relation from the mode $J^3_{-1}$. We start with following
\bea 
&&\s^+(w)J^3_{-1}|0^-_R\rangle^{(1)}|0^-_R\rangle^{(2)}\cr 
&=&\!\!-\s^+(w){1\over2}(d^{++(1)}_{0}d^{--(1)}_{-1}  - d^{+-(1)}_{0}d^{-+(1)}_{-1}  + d^{++(2)}_{0}d^{--(2)}_{-1} - d^{+-(2)}_{0}d^{-+(2)}_{-1})|0^-_R\rangle^{(1)}|0^-_R\rangle^{(2)}\nn
\eea
Bringing the modes through the twist, two fermionic modes can contract with each other and leave terms which contain no modes. Using the contraction (\ref{Cfull}) we obtain
\bea \label{con J-1 0}
-(C^{11}[0,1]  + C^{22}[0,1]  )|0^{2-}_R\rangle = 2C^2e^{-w}|0^{2-}_R\rangle
\eea 
Again looking at the state
\bea 
\s^+(w)J^3_{-1}|0^-_R\rangle^{(1)}|0^-_R\rangle^{(2)}=(J^3_{-1}\s^+(w) - [J^3_{-1},\s^+(w)])|0^-_R\rangle^{(1)}|0^-_R\rangle^{(2)}
\eea
Inserting the commutator (\ref{Jm1}) into the above gives
\bea
\s^+(w)J^3_{-1}|0^-_R\rangle^{(1)}|0^-_R\rangle^{(2)}&=&\bigg(J^3_{-1} - {1\over2}e^{-w}\bigg)\s^+(w)|0^-_R\rangle^{(1)}|0^-_R\rangle^{(2)}\cr
&=&\bigg(J^3_{-1} - {1\over2}e^{-w}\bigg)|\chi\rangle
\eea
Expanding and only keeping terms which do not contain any modes we get 
\bea\label{con J-1}
-{1\over2}e^{-w}|0^{2-}_R\rangle
\eea
Comparing (\ref{con J-1 0}) and (\ref{con J-1}),  we obtain 
\bea\label{Const sq}
C^2=-{1\over4}
\eea
which gives
\bea\label{Const}
C=\pm {i\over2}
\eea
This is the constant appearing in the propagation (\ref{full f-}) and (\ref{full f+}). The two choices of sign correspond to the two different conventions of labeling the copies.

\section{Relations from supersymmetry}\label{susy}

In the previous sections we derived the effect of the twist operator for fermionic modes. A similar method can be applied to bosonic modes. See \cite{Guo:2022sos} for a toy model. Since the D1D5 CFT has supersymmetry, we will use this to derive the effects for the bosonic modes from the effects of the fermionic modes.

Let us first find the propagation. We start with
\be
G^{+}_{\dot A,0}d^{-B(i)}_{-n}|0^-_R\rangle^{(1)}|0^-_R\rangle^{(2)} = -i \epsilon^{AB} \alpha^{(i)}_{A \dot A,-n}|0^-_R\rangle^{(1)}|0^-_R\rangle^{(2)}
\ee
where we have used the commutator in (\ref{commutations}) and the fact that $G^{+}_{\dot A,0}$ annihilates any Ramond ground state
\be
G^{+}_{\dot A,0}|0^-_R\rangle^{(1)}|0^-_R\rangle^{(2)} = 0
\ee
Applying the twist operator, we obtain the relation
\be\label{Gd}
\s^+(w)G^{+}_{\dot A,0}d^{-B(i)}_{-n}|0^-_R\rangle^{(1)}|0^-_R\rangle^{(2)} = -i \epsilon^{AB} \s^+(w)\alpha^{(i)}_{A \dot A,-n}|0^-_R\rangle^{(1)}|0^-_R\rangle^{(2)}
\ee
Commuting the mode $G^{+}_{\dot A,0}$ through the twist operator using 
\be\label{Gsigma}
[G^{+}_{\dot A,0},\s^+(w)]=0
\ee
the LHS becomes
\bea\label{GdL}
G^{+}_{\dot A,0}\s^+(w)d^{-B(i)}_{-n}|0^-_R\rangle^{(1)}|0^-_R\rangle^{(2)}
&=& G^{+}_{\dot A,0}\sum_{p>0} f^{-}_{i}[-n,-p] d^{-B}_{-p}|0^{2-}_R\rangle \nn
&=& -i \epsilon^{AB}\sum_{p>0} f^{-}_{i}[-n,-p] \alpha_{A\dot A,-p}|0^{2-}_R\rangle
\eea
Using the ansatz (\ref{single mode}), the RHS of (\ref{Gd}) becomes
\be\label{GdR}
-i \epsilon^{AB} \sum_{p>0} f_{i}[-n,-p]\alpha_{A \dot A,-p}|0^-_R\rangle^{(1)}|0^-_R\rangle^{(2)}
\ee
Comparing (\ref{GdL}) and (\ref{GdR}), we find
\be\label{f boson}
f_{i}[-n,-p] = f^-_{i}[-n,-p]
\ee

Let us now find the relation for pair creation. We start with
\be
G^{+}_{\dot A,0}|0^-_R\rangle^{(1)}|0^-_R\rangle^{(2)} = 0
\ee
Applying the twist operator we obtain
\be
\sigma^+(w)G^{+}_{\dot A,0}|0^-_R\rangle^{(1)}|0^-_R\rangle^{(2)} = 0
\ee
Commuting the mode $G^{+}_{\dot A,0}$ through the twist operator using (\ref{Gsigma}) and taking a specific choice $\dot A =+$ we find
\bea
0&=&G^{+}_{+,0}\sigma^+(w)|0^-_R\rangle^{(1)}|0^-_R\rangle^{(2)}\nn
&=&G^{+}_{+,0}\sum_{m,n>0}\big(\gamma^B_{mn}(-\a_{++,-m}\a_{--,-n} + \a_{+-,-m}\a_{-+,-n})\nn
&&\hspace{2cm}+ \gamma^F_{mn}(d^{++}_{-m}d^{--}_{-n} - d^{+-}_{-m}d^{-+}_{-n}) \big)|0^{2-}_R\rangle+\dots\nn
&=&\sum_{m,n>0}\big(\gamma^B_{mn}  (-i n\a_{++,-m}d^{++}_{-n} - im d^{+-}_{-m}\a_{-+,-n})\nn
&&\hspace{2cm}+ \gamma^F_{mn}(-id^{++}_{-m}\a_{++,-n} - id^{+-}_{-m}\a_{-+,-n}) \big)|0^{2-}_R\rangle+\dots
\eea
which implies
\be\label{boson gamma}
\gamma^B_{mn}=-\frac{\gamma^F_{mn}}{m}
\ee

Let us now find the contraction. We start with
\bea\label{sc1}
\s^+(w)\alpha^{(i)}_{++,-n}\alpha^{(j)}_{--,-m}|0^-_R\rangle^{(1)}|0^-_R\rangle^{(2)} 
= - C_B^{ij}[n,m]|0^{2-}_R\rangle + \dots
\eea
where we have kept only the term without any mode. This term can also be computed as follows
\bea\label{sc2}
&&\s^+(w)\alpha^{(i)}_{++,-n}\alpha^{(j)}_{--,-m}|0^-_R\rangle^{(1)}|0^-_R\rangle^{(2)} \nn
&=& -i \s^+(w)\{G^+_{+,0},d^{--(i)}_{-n}\}\alpha^{(j)}_{--,-m}|0^-_R\rangle^{(1)}|0^-_R\rangle^{(2)} \nn
&=& -i \s^+(w)G^+_{+,0}d^{--(i)}_{-n}\alpha^{(j)}_{--,-m}|0^-_R\rangle^{(1)}|0^-_R\rangle^{(2)}
-i \s^+(w)d^{--(i)}_{-n}G^+_{+,0}\alpha^{(j)}_{--,-m}|0^-_R\rangle^{(1)}|0^-_R\rangle^{(2)}\nn
&=& -iG^+_{+,0} \s^+(w)d^{--(i)}_{-n}\alpha^{(j)}_{--,-m}|0^-_R\rangle^{(1)}|0^-_R\rangle^{(2)}
-i \s^+(w)d^{--(i)}_{-n}(im)d^{++(j)}_{-m}|0^-_R\rangle^{(1)}|0^-_R\rangle^{(2)}\nn
&=& -m C^{ij}_F[m,n]|0^{2-}_R\rangle + \dots
\eea
To get the fourth line, we commute $G^+_{+,0}$ through $\s^+(w)$ in the first term and commute $G^+_{+,0}$ with $\alpha_{--,-m}$ in the second term. To obtain the last line, notice that the first term in the fourth line does not contain a term without any mode. We have also used the fact that $C^{ij}_F[n_1,n_2]$ is symmetric under the interchange of the copy labels $i$ and $j$. For the second term we have kept only terms without any modes.
Comparing (\ref{sc1}) and (\ref{sc2}) we obtain
\be\label{boson C}
C^{ij}_B[n,m] = m C^{ij}_F[m,n]
\ee
Thus by using (\ref{f boson}), (\ref{boson gamma}) and (\ref{boson C}) we get the effect of the twist operator for bosons from fermions. The result is summarized in the next section.

\section{Solutions}\label{sec solution}

In this section we record the full expressions for pair creation, propagation, and contraction for both bosons and fermions. Our results agree with the expressions computed in \cite{Avery:2010er} and \cite{Avery:2010hs} which use the covering map method.

\subsection{Fermions}

\subsubsection*{Pair creation}

The expression for pair creation is given by 
\bea\label{gamma} 
\gamma^F_{m+{1\over2},n+{1\over2}} = -{e^{(m+n+1)w}\Gamma[{3\over2}+m]\Gamma[{3\over2}+n]\over(2n+1)\pi(m+n+1)\Gamma[m+1]\Gamma[n+1]} ~~~ ~~~m,n\geq0
\eea
where $m$ and $n$ are integers.

\subsubsection*{Propagation}

Choosing the value of the constant $C$ with a negative sign in (\ref{Const}) and inserting this into (\ref{full f-}) and (\ref{full f+}) we obtain the expressions
\bea
\!\!f_j^-[-n,-(m+1/2)]&=&-(-1)^{j}{i\Gamma[{1\over2} + n] \Gamma[{1\over2} + m]\over\pi\Gamma[n] \Gamma[1 + m]}{e^{(m - n + 1/2) w}\over2 m - 2 n + 1},~~~ ~~~n>0, m\geq 0
\cr 
f^+_j[-n,-(m+1/2)] &=& -(-1)^{j}\frac{i\Gamma[\frac{1}{2}+n]}{\pi\Gamma[1+n]}
\frac{\Gamma[\frac{3}{2}+m]}{\Gamma[1+m]}\frac{e^{(m-n+1/2)w}}{2m-2n+1},~~~ ~~~n\geq0, m\geq 0\nn
\eea
where $m$ and $n$ are integers. As shown in appendix \ref{app global}, when the initial and final mode are equal, we have
\bea
f^{\pm}_i[-n,-n]={1\over2}
\eea
where $n$ is a strictly positive integer for $f^-_i$ and a non-negative integer for $f^+_i$.
\subsubsection*{Contraction}

Inserting the value $C$ chosen above into (\ref{Cfull}) (Since $C^2$ appears in the contraction term, you get the same solution when taking either sign) we obtain
\bea\label{Cfull}
C^{ij}_F[n_1,n_2] =(-1)^{i+j}{\Gamma[{1\over2}+n_1]\Gamma[{1\over2}+n_2]\over 2\pi n_1\Gamma[n_1]\Gamma[n_2]}{e^{-(n_1+n_2)w}\over n_1+n_2}, ~~~ ~~~ n_1\geq 0, n_2\geq 1
\eea
and therefore
\bea 
C^{ij,\a A\beta B} &=& \e^{\a\beta}\e^{AB}C^{ij}_F[n_1,n_2]\cr 
&=& \e^{\a\beta}\e^{AB}(-1)^{i+j}{\Gamma[{1\over2}+n_1]\Gamma[{1\over2}+n_2]\over 2\pi n_1\Gamma[n_1]\Gamma[n_2]}{e^{-(n_1+n_2)w}\over n_1+n_2}, ~~~ ~~~ n_1\geq 0, n_2\geq 1
\eea
\subsection{Bosons}
Using the relations derived in section \ref{susy} we record the expressions for the bosonic quantities.

\subsubsection*{Pair creation}
For pair creation we use relation (\ref{boson gamma}) and obtain
\bea 
\gamma^B_{m+{1\over2},n+{1\over2}} &=& {2e^{(m+n+1)w}\Gamma[{3\over2}+m]\Gamma[{3\over2}+n]\over(2m+1)(2n+1)\pi(m+n+1)\Gamma[m+1]\Gamma[n+1]}, ~~~ ~~~m,n\geq0
\eea

\subsubsection*{Propagation}
For propagation we use relation (\ref{f boson}). This gives 
\bea
\!\!f_j[-n,-(m+1/2)]=-(-1)^{j}{i\Gamma[{1\over2} + n] \Gamma[{1\over2} + m]\over\pi\Gamma[n] \Gamma[1 + m]}{e^{(m - n + 1/2) w}\over2 m - 2 n + 1},~~~n>0, m\geq 0
\eea
where $m,n$ are integers. As shown in appendix \ref{app global}, when the initial and final mode are equal, we have
\bea
f_i[-n,-n]={1\over2}
\eea
where $n$ is a positive integer.
\subsubsection*{Contraction}
For contraction, using relation (\ref{boson C}), we obtain 
\bea
C^{ij}_B[n_1,n_2] =(-1)^{i+j}{\Gamma[{1\over2}+n_1]\Gamma[{1\over2}+n_2]\over 2\pi \Gamma[n_1]\Gamma[n_2]}{e^{-(n_1+n_2)w}\over n_1+n_2}, ~~~ ~~~ n_1\geq 1, n_2\geq 1
\eea
and therefore
\bea
C^{ij}_{A\dot AC\dot C}[n_1,n_2]&=&\e_{AC}\e_{\dot A\dot C}C^{ij}_B[n_1,n_2]\cr 
&=& \e_{AC}\e_{\dot A\dot C}(-1)^{i+j}{\Gamma[{1\over2}+n_1]\Gamma[{1\over2}+n_2]\over 2\pi \Gamma[n_1]\Gamma[n_2]}{e^{-(n_1+n_2)w}\over n_1+n_2}, ~~~ ~~~ n_1\geq 1, n_2\geq 1\nn
\eea
where we have used the symmetry between the exchange of bosonic modes.

\section{Discussion}\label{discussion}

In this paper we have used the bootstrap method developed in \cite{Guo:2022sos} to compute the effects of the twist operator in the D1D5 CFT. The majority of the paper has focused on computing the effects involving fermionic modes. In section \ref{susy} the bosonic quantities are then derived from the fermionic ones using supersymmetry relations. The effects involving the fermions are fully captured by three quantites: pair creation $\gamma^F_{mn}$, propagation $f^{\pm}_i[-n,-p]$, and contraction $C_F^{ij}[n_1,n_2]$. Using the weak Bogoliubov ansatz and superconformal symmetry we were able to derive expressions for these quantities. 
We were able to compute the expression for pair creation $\gamma^F_{mn}$ using the generator $L_{-1} + J^3_{-1}$. Knowing the expression for pair creation, we then computed the expression for propagation $f^-_i[-n,-p]$ using the generator $J^3_1$ and then the expression for propagation $f^+_i[-n,-p]$ from $f^-_i[-n,-p]$ using the generator $J^+_1$.  
Furthermore, using the generator $J^3_1$, we were able to compute the expression for contraction $C^{ij}_F[n_1,n_2]$ by knowing $f^+_i[-n,-p]$ and $f^-_i[-n,-p]$. 
From the fermionic quantities, we then used the supersymmetric generator $G^+_{\dot A,0}$ to derive the expressions for bosons. 

All results in this paper were derived for a single twist operator. A major goal of this program is to eventually compute effects and correlators which contain an arbitrary number of twist operators. This would help to obtain a CFT description of certain quantities in the supergravity regime. Under some approximations, it seems very promising to compute some coefficients of the effect for an arbitrary number of twist operators. 
Here, we also considered only two singly wound copies in the initial state, twisting them into a doubly wound copy in the final state. We would like to use the bootstrap approach to derive effects of the twist operator with multiwound copies in the initial state. There has been much recent work in developing an exact correspondence between the string worldsheet with NSNS flux and the orbifold CFT. The bootstrap techniques developed in this and the previous paper may be helpful in better understanding this correspondence. We hope to return to this in a future work.

\section*{Acknowledgements}

We would like to thank Soumangsu Chakraborty, Nicolas Kovensky, Samir Mathur and Hynek Paul for helpful discussions.
The work of B.G. and S.D.H. is supported by ERC Grant 787320 - QBH Structure.

\appendix
\section{$\mathcal{N}=4$ superconformal algebra}
In this appendix we record the $\mathcal{N}=4$ superconformal algebra \cite{Schwimmer:1986mf,Sevrin:1988ew} for $c=6k$. We start with the basic commutation relations which are composed of 4 bosons and 4 fermions
\bea\label{basic com}
[\a_{A\dot{A},m},\a_{B\dot{B},n}] &=& - k m\e_{A\dot{A}}\e_{B\dot{B}}\delta_{m+n,0}\cr
\lbrace d^{\alpha A}_r , d^{\beta B}_s\rbrace  &=&- k \e^{\alpha\beta}\e^{AB}\delta_{r+s,0}
\eea
The superconformal algebra is composed of Virasoro generators $L_m$, SU(2) current generators $J^a$ which form a Kac-Moody algebra where $a=1,2,3$, and superconformal generators $G^{\a}_{\dot A}$ where $\a = +,-$ and $\dot A=+,-$. For $J^1$ and $J^2$ we define $J^{\pm}$.
\bea
J^\pm_n=J^1_n \pm i J^2_n
\eea
The current-current commutation relations are given by
\bea\label{com}
[L_m,L_n] &=& {c\over12}m(m^2-1)\delta_{m+n,0}+ (m-n)L_{m+n}\cr
[J^a_{m},J^b_{n}] &=&{c\over12}m\delta^{ab}\delta_{m+n,0} +  i\e^{ab}_{\,\,\,\,c}J^c_{m+n}\cr
[L_{m},J^a_n]&=& -nJ^a_{m+n}\cr
[L_{m},G^{\a}_{\dot{A},r}] &=& ({m\over2}  -r)G^{\a}_{\dot{A},m+r}\cr
[J^a_{m},G^{\a}_{\dot{A},r}] &=&{1\over2}(\s^{aT})^{\a}_{\beta} G^{\beta}_{\dot{A},m+r}\cr
\lbrace G^{\a}_{\dot{A},r} , G^{\beta}_{\dot{B},s} \rbrace&=&  \e_{\dot{A}\dot{B}}\bigg[\e^{\a\beta}{c\over6}(r^2-{1\over4})\delta_{r+s,0}  + (\s^{aT})^{\a}_{\g}\e^{\g\beta}(r-s)J^a_{r+s}  + \e^{\a\beta}L_{r+s}  \bigg]
\cr
[J^3_m , J^{+}_n] &=& J^{+}_{m+n},\qquad\qquad [J^3_m , J^{-}_n] ~=~ -J^{-}_{m+n}\cr
[J^+_{m},J^-_{n}]&=&{c\over6}m\delta_{m+n,0} + 2J^3_{m+n}\cr
[J^{3}_{m},G^{\pm}_{\dot{A},r}]  &=& \pm \frac{1}{2}G^{\pm}_{\dot{A},m+r} \cr
[J^{+}_{m},G^{+}_{\dot{A},r}]  &=& 0 ,\qquad\qquad ~~~[J^{-}_{m},G^{+}_{\dot{A},r}]  ~=~ G^{-}_{\dot{A},m+r}\cr
[J^{+}_{m},G^{-}_{\dot{A},r}]  &=&G^{+}_{\dot{A},m+r},\qquad ~[J^{-}_{m},G^{-}_{\dot{A},r}]  ~=~ 0 
\eea 

Using the free field realization, we can expand the generators in terms of bosonic and fermionic modes
\bea\label{g}
J^a_m &=& {1\over 4 k}\sum_{r}\epsilon_{AB}d^ {\g B}_r\epsilon_{\alpha\gamma}(\s^{aT})^{\a}_{\beta}d^ {\beta A}_{m-r},\qquad a=1,2,3\cr
J^3_m &=&  - {1\over 2 k}\sum_{r} d^ {+ +}_{r}d^ {- -}_{m-r} - {1\over 2 k}\sum_{r}d^ {- +}_r d^ {+ -}_{m-r}\cr
J^{+}_m&=& {1\over k}\sum_{r}d^ {+ +}_rd^ {+ -}_{m-r} ,\qquad J^{-}_m= {1\over k} \sum_{r}d^ {--}_rd^ {- +}_{m-r}\cr
G^{\a}_{\dot{A},r} &=& - {i\over k} \sum_{n}d^ {\a A}_{r-n} \a_{A\dot{A},n}\cr
L_m&=& -{1\over 2 k}\sum_{n} \e^{AB}\e^{\dot A \dot B}\a_{A\dot{A},n}\a_{B\dot{B},m-n}- {1\over 2 k}\sum_{r}(m-r+{1\over2})\epsilon_{\alpha\beta}\epsilon_{AB}d^ {\a A}_r d^ {\beta B}_{m-r}
\eea

Below we record the commutation relations between the generators and bosonic and fermionic modes
\bea\label{commutations}
[L_m,\a_{A\dot{A},n}] &=&-n\a_{A\dot{A},m+n} \cr
[L_m ,d^{\a A}_r] &=&-({m\over2}+r)d^{\a A}_{m+r}\cr
\lbrace G^{\a}_{\dot{A},r} ,  d^{\beta B}_{s} \rbrace&=&i\e^{\a\beta}\e^{AB}\a_{A\dot{A},r+s}\cr
[G^{\a}_{\dot{A},r} , \a_{B \dot{B},m}]&=&  -im\e_{AB}\e_{\dot{A}\dot{B}}d^{\a A}_{r+m}\cr
[J^a_m,d^{\a A}_r] &=&{1\over 2}(\s^{Ta})^{\a}_{\beta}d^{\beta A}_{m+r}
\eea
\bea
[J^{3}_m,d^{\pm A}_r] &=& \pm \frac{1}{2} d^{\pm A}_{m+r}\cr
[J^{+}_m,d^{+ A}_r] &=& 0,\qquad~~~~~ [J^{-}_m,d^{+ A}_r] ~=~ d^{-A}_{m+r}\cr
[J^{+}_m,d^{- A}_r] &=& d^{+A}_{m+r},\qquad [J^{-}_m,d^{- A}_r] ~=~ 0
\eea

\section{Global modes}\label{app global}

In this appendix we will show (\ref{f hi}) and (\ref{g hi}). Let us consider, for an integer $n$, 
\bea\label{g f}
&&d^{\alpha A}_n \s^+(w) - \s^+(w) (d^{\alpha A(1)}_n+d^{\alpha A(2)}_n) \nn
&=& {1\over2\pi i}\oint_{C_{w}^{(4\pi)}}dw' e^{{n}w'} \psi^{\a A}(w') \s^+(w)\nn
&=& e^{nw}{1\over2\pi i}\oint_{C_{w}^{(4\pi)}}dw' e^{n(w'-w)} \psi^{\a A}(w') \s^+(w) \nn
&=& e^{nw}{1\over2\pi i}\oint_{C_{w}^{(4\pi)}}dw' (1+n(w'-w)+\dots) \psi^{\a A}(w') \s^+(w) \nn
&=& e^{nw}(d^{\alpha A(w)}_{1/2}+n d^{\alpha A(w)}_{3/2} + \dots) \s^+(w)
\eea
To get the second line, we join the two contours, one before and one after the twist operator $\s(w)$, into a single contour around the twist operator. We denote the modes centered around $w$ by a superscript $(w)$, e.g. $d^{\alpha A(w)}_{1/2}$. Since there is no local operator with dimension $h = 0$ and a nonzero $A$ charge. Furthermore, there is no operator with $h<0$, we have
\be
d^{\alpha A(w)}_{m+1/2}\s^+(w)=0,~~~~~m\geq 0
\ee
where $m$ is an integer. Thus (\ref{g f}) implies
\be\label{g f1}
d^{\alpha A}_n \s^+(w) - \s^+(w) (d^{\alpha A(1)}_n+d^{\alpha A(2)}_n)=0
\ee
Similarly, we have for bosons
\bea\label{g b}
\alpha_{A \dot A,n} \s^+(w) - \s^+(w) (\alpha^{(1)}_{A \dot A,n}+\alpha^{(2)}_{A \dot A,n})
= e^{nw}(\alpha^{(w)}_{A \dot A,0}+n \alpha^{(w)}_{A \dot A,1} + \dots) \s^+(w) = 0
\eea
where we have used the fact that
\be
\alpha^{(w)}_{A \dot A,n}\s^+(w) =0, ~~~~n\geq 0
\ee

Consider the following state where $n$ is a positive integer if $\beta=-$ and a non-negative integer if $\beta=+$. 
Using (\ref{g f1}), we get
\bea
\langle 0^{2-}_R| d^{\alpha A}_{m}\s^+(w)d^{\beta B(i)}_{-n}|0^-_R\rangle^{(1)}|0^-_R\rangle^{(2)} 
&=&\langle 0^{2-}_R| \s^+(w)(d^{\alpha A(1)}_{m}+d^{\alpha A(2)}_{m})d^{\beta B(i)}_{-n}|0^-_R\rangle^{(1)}|0^-_R\rangle^{(2)}\nn
&=& - \epsilon^{\alpha \beta}\epsilon^{A B}\delta_{m,n}
\eea
which is nonzero only if $m$ and $n$ are equal. Because
\be
\langle 0^{2-}_R| d^{\alpha A}_{m}d^{\beta B}_{-n}|0^{2-}_R\rangle = - 2\epsilon^{\alpha \beta}\epsilon^{A B}\delta_{m,n}
\ee
we find 
\be
\s^+(w)d^{\beta B(i)}_{-n}|0^-_R\rangle^{(1)}|0^-_R\rangle^{(2)}
=\left(\frac{1}{2}d^{\beta B}_{-n}+\text{half integer modes}\right)|0^{2-}_R\rangle
\ee

Similarly for bosons, consider the following state where $n$ is a positive integer. Using (\ref{g b}), we get
\bea
\langle 0^{2-}_R| \alpha_{A\dot A,m}\s^+(w)\alpha^{(i)}_{B\dot B,-n}|0^-_R\rangle^{(1)}|0^-_R\rangle^{(2)} 
\!\!&=&\!\!\langle 0^{2-}_R| \s^+(w)(\alpha^{(1)}_{A\dot A,m}+\alpha^{(2)}_{A\dot A,m})\alpha^{(i)}_{B\dot B,-n}|0^-_R\rangle^{(1)}|0^-_R\rangle^{(2)}\nn
\!\!&=&\!\! - m\epsilon_{AB}\epsilon_{\dot A \dot B}\delta_{m,n}
\eea
which is nonzero only if $m$ and $n$ are equal. Because
\be
\langle 0^{2-}_R| \alpha_{A\dot A,m}\alpha_{B\dot B,-n}|0^{2-}_R\rangle = -2 m\epsilon_{AB}\epsilon_{\dot A \dot B}\delta_{m,n}
\ee
we find 
\be
\s^+(w)\alpha^{(i)}_{B\dot B,-n}|0^-_R\rangle^{(1)}|0^-_R\rangle^{(2)}
=\left(\frac{1}{2}\alpha_{B\dot B,-n}+\text{half integer modes}\right)|0^{2-}_R\rangle
\ee
Thus we have shown (\ref{f hi}).

To show (\ref{g hi}), consider the following state where $m$ is a positive integer for $\alpha=+$ and a non-negative integer for $\alpha=-$
\be
d^{\alpha A}_{m}|\chi\rangle = d^{\alpha A}_{m} \s^+(w) |0^-_R\rangle^{(1)}|0^-_R\rangle^{(2)}
= \s^+(w)(d^{\alpha A(1)}_{m}+d^{\alpha A(2)}_{m}) |0^-_R\rangle^{(1)}|0^-_R\rangle^{(2)} = 0
\ee
Similarly for bosons, consider the following state where $m$ is a positive integer
\be
\alpha_{A\dot A,m}|\chi\rangle = \alpha_{A\dot A,m} \s^+(w) |0^-_R\rangle^{(1)}|0^-_R\rangle^{(2)}
= \s^+(w)(\alpha^{(1)}_{A\dot A,m}+\alpha^{(2)}_{A\dot A,m}) |0^-_R\rangle^{(1)}|0^-_R\rangle^{(2)} = 0
\ee
Thus the state $\chi$ from pair creation should not have any integer mode excitations, which is stated in (\ref{g hi}).

\section{Propagation: higher modes}\label{app f-}

In section \ref{propagation}, we have derived $f_i^-[-1,-(p+1/2)]$.
In this appendix we derive the expression for $f_i^-[-n,-(p+1/2)]$ for any $n>0$.
We start with the following state
\bea
d^{--(i)}_{-n} |0^-_R\rangle^{(1)}|0^-_R\rangle^{(2)} =  {1\over (n-1)!}(L_{-1}+J^3_{-1})^{n-1} d^{--(i)}_{-1} |0^-_R\rangle^{(1)}|0^-_R\rangle^{(2)}
\eea
where $n>0$.
Applying the twist operator gives
\bea\label{sigma dpp n}
\sigma^+(w) d^{--(i)}_{-n} |0^-_R\rangle^{(1)}|0^-_R\rangle^{(2)}
=\frac{1}{(n-1)!}\sigma^+(w)  (L_{-1}+J^3_{-1})^{n-1} d^{--(i)}_{-1} |0^-_R\rangle^{(1)}|0^-_R\rangle^{(2)}
\eea
Commuting $(L_{1}+J^3_{-1})^{n-1}$ through the twist and using (\ref{Lm1}) and (\ref{Jm1}) we find the relation
\bea\label{fmderivation}
&&{1\over (n-1)!}\sigma^+(w) (L_{-1}+J^{3}_{-1})^{n-1} d^{--(i)}_{-1} |0^-_R\rangle^{(1)}|0^-_R\rangle^{(2)}\cr
&=&
{1\over (n-1)!}\big[L_{-1}+J^{3}_{-1} -e^{-w}\partial\big]^{n-1}\sigma^+(w)d^{--(i)}_{-1} |0^-_R\rangle^{(1)}|0^-_R\rangle^{(2)}\cr
&=&{1\over (n-1)!}\sum_{k=0}^{n-1} {}^{n-1}C_k (L_{-1} + J^3_{-1})^{k} \big(-e^{-w}\partial\big)^{n-k-1} \sigma^+(w)d^{--(i)}_{-1} |0^-_R\rangle^{(1)}|0^-_R\rangle^{(2)}\cr
&=&
{1\over (n-1)!}\sum_{k=0}^{n-1} {}^{n-1}C_k (L_{-1} + J^{3}_{-1} )^{k} \big(-e^{-w}\partial\big)^{n-k-1}\nn
&&\hspace{2.5cm}\bigg({1\over2}d^{--}_{-1} + \sum_{p\geq 0}f^-_i[-1,-(p+1/2)]d^{--}_{-(p + 1/2)}\bigg)|0^{2-}_R\rangle
+\dots
\eea
where we have kept terms with only one $d^{--}$ mode.
We note that ${}^{n-1}C_k$ is the binomial coefficient
\bea
{}^{n-1}C_k={(n-1)!\over k!(n-k-1)!}
\eea
We want to compare the expression (\ref{fmderivation}) to the left hand side of (\ref{sigma dpp n}) containing only one $d^{--}$ which is given by
\bea\label{lhs}
\sigma^+(w) d^{--(1)}_{-n} |0^-_R\rangle^{(1)}|0^-_R\rangle^{(2)}=\bigg({1\over2}d^{--}_{-n} + \sum_{p\geq0}f^-_i[-n,-(p+1/2)]d^{--}_{-(p + 1/2)} + \ldots\bigg)|0^{2-}_R\rangle
\eea
To match both (\ref{fmderivation}) and (\ref{lhs}) we pick the $m$'th term out of the sum over half integer modes, 
which is the state
\be
d^{--}_{-(m+1/2)}|0^{2-}_R\rangle
\ee
 For the expression (\ref{fmderivation}), this requires us to relabel the indices taking
\be
p+k=m
\ee
This is because each $L_{-1} + J^3_{-1}$ increases the dimension by 1 unit. The limits of the sum are determined by
\bea
&&k\geq 0\implies m\geq p\cr
&& p\geq 0\text{ and }k\leq n-1\implies p \geq \max(m-(n-1),0)
\eea
Therefore the relevant term in (\ref{fmderivation}) is
\bea\label{rhs two}
&&{1\over (n-1)!}\sum_{p=\text{max}(m-(n-1),0)}^{m-1} {}^{n-1}C_k (L_{-1} + J^{3}_{-1} )^{m-p} \big(-e^{-w}\partial\big)^{n-(m-p)-1}\nn
&&\hspace{5cm}f^-_i[-1,-(p+1/2)]d^{--}_{-(p + 1/2)}|0^{2-}_R\rangle
\eea
Let's determine the action of $L_{-1} + J^3_{-1}$. We have
\bea\label{LJ}
(L_{-1}+J^3_{-1})d^{--}_{-(p + 1/2)} |0^{2-}_R\rangle 
 &=& \bigg(p  + {1\over2}\bigg) d^{++}_{-(p+3/2) }|0^{2-}_R\rangle +\dots
\eea 
 This implies that
\bea
&&(L_{-1} + J^3_{-1} )^{n'}d^{--}_{-(p + 1/2)} |0^{2-}_R\rangle\cr 
&&=\bigg(p + {1\over2}\bigg) \bigg(p+{3\over2}\bigg)\dots \bigg(p+n'-{1\over2}\bigg)d^{--}_{-(p+n'+1/2) }|0^{2-}_R\rangle+\dots\cr
 &&= \frac{(p+n'-{1\over2})!}{(p-{1\over2})!}d^{--}_{-(p+n'+1/2) } |0^{2-}_R\rangle +\dots
\eea
where we have kept the terms with only one $d^{--}$ mode. Let us also determine the action of $e^{-w}\partial$. To do so we notice that $f^-_i[1,-(p+1/2)]\propto e^{(p-\frac{1}{2})w}$ so we have
\bea
(e^{-w}\partial)f^-_i[-1,-(p+1/2)] &=&f^-_{i}[-1,-(p+1/2)]_{w=0}(e^{-w}\partial)e^{(p-\frac{1}{2})w}\cr
&=& f^-_{i}[-1,-(p+1/2)]_{w=0}e^{-w} \bigg(p-{1\over2}\bigg) e^{(p-1-\frac{1}{2})w}
\eea
and therefore
\bea\label{partial}
&&(e^{-w}\partial)^{n''}f^-_{i}[-1,-(p+1/2)]\cr 
&& =   f^-_{i}[-1,-(p+1/2)]_{w=0} \bigg(p-{1\over2}\bigg)\bigg(p-{3\over2}\bigg)\dots\bigg(p-(n''-1)-{1\over2}\bigg) e^{(p-n''-\frac{1}{2})w}\cr
&&={(p-{1\over2})!\over(p-n''-{1\over2})!}e^{(p-n''-\frac{1}{2})w}f^-_i[-1,-(p+1/2)]_{w=0}
\eea
Looking at (\ref{rhs two}) we set $n'=m-p$ in (\ref{LJ}) and $n''= n - (m-p)-1$ in (\ref{partial}). This gives the expressions
\bea
(L_{-1}+J^3_{-1})^{m-p} d^{--}_{-(p + 1/2)}  |0^-_R\rangle &=&\frac{(m-{1\over2})!}{(p-{1\over2})!}d^{--}_{-(m + 1/2)}  |0^-_R\rangle \cr
(e^{-w}\partial)^{n - (m-p)-1}f^-_i[-1,-(p+1/2)]&=&{(p-{1\over2})!\over(m-n+{1\over2})!}e^{(m -n + {1\over2})w}f^-_i[-1,-(p+1/2)]_{w=0}\nn
\eea
Finally (\ref{fmderivation}) becomes
\bea\label{rhs three}
&&{1\over(n-1)!}\sigma^+(w) (L_{-1}  + J^3_{-1}  )^{n-1} d^{--(i)}_{-1} |0^-_R\rangle^{(1)}|0^-_R\rangle^{(2)}\cr
&&\quad = e^{(m-n+\frac{1}{2})w}{1\over (n-1)!}\sum_{p= \max(m-(n-1),0)}^{m} {}^{n-1}C_{m-p} \frac{(m-{1\over2})!}{(m-n+{1\over2})!} (-1)^{n-(m-p)-1}\cr
&& \hspace{5cm}  f^-_i[-1,-(p+1/2)]_{w=0}d^{--}_{-(m + 1/2)}|0^-_R\rangle + \dots
\eea
Comparing (\ref{rhs three}) with the term $d^{--}_{-(m+1/2)}|0^{2-}_R\rangle$ in (\ref{lhs}) we obtain the relation
\bea\label{fullfm}
&&f^-_i[-n,-(m+1/2)]\cr
&&\quad=e^{(m-n+\frac{1}{2})w}\frac{1}{(n-1)!}\frac{(m-{1\over2})!}{(m-n+{1\over2})!}\nn
&&\hspace{1cm}\sum_{p= \max(m-(n-1),0)}^{m} {}^{n-1}C_{m-p} (-1)^{n-(m-p)-1}f^-_i[-1,-(p+1/2)]_{w=0}
\eea
Inserting $f^-_i[-1,-(p+1/2)]$ (\ref{f-1}) into the above expression yields
\bea 
f^-_i[-n,-(m+1/2)]=C(-1)^{i}{2\Gamma[{1\over2} + n] \Gamma[{1\over2} + m ]\over\pi\Gamma[n] \Gamma[1 + m ]}{e^{(m  - n + 1/2) w}\over2 m  - 2 n + 1}
\eea

\section{Contraction: higher modes}\label{app C}

In section \ref{C} the contraction $C^{ij}_F[n,1]$ for $n\geq 0$ was found.
In this appendix we compute the full $C^{ij}_F[n_1,n_2]$ for $n\geq 0$ and $m\geq 1$. We start with the state
\bea\label{Cn1n2}
|\psi\rangle&=&\s^+(w)d^{++(i)}_{-n_1}d^{--(j)}_{-n_2}|0^-_R\rangle^{(1)}|0^-_R\rangle^{(2)}\cr 
&=& C_F^{ij}[n_1,n_2]|0^{2-}_R\rangle + \ldots
\eea
where in the second line we have kept only the unexcited $|0^{2-}_R\rangle$.
Returning back to the original expression of $|\psi\rangle$ 
and using $L_{-1} + J^{3}_{-1}$ we can write this as
\bea \label{C2}
|\psi\rangle=\s^+_2(w){1\over (n_2-1)!}d^{++(i)}_{-n_1}(L_{-1} + J^{3}_{-1})^{n_2-1}d^{--(j)}_{-1}|0^-_R\rangle^{(1)}|0^-_R\rangle^{(2)} 
\eea
To compute the above expression we notice that
\bea
d^{++(i)}_{-n_1}(L_{-1} + J^{3}_{-1})^{n_2-1} = \big(L_{-1} + J^3_1-(L_{-1} + J^3_1)\circ\big)^{n_2-1}d^{++(i)}_{-n_1}
\eea
where for some operator $O_{-n}$
\bea
(L_{-1} + J^3_1)\circ O_{-n}\equiv [ L_{-1} + J^3_1, O_{-n}]
\eea
Thus we have
\bea
&&d^{++(i)}_{-n_1}(L_{-1} + J^{3}_{-1})^{n_2-1}\nn
&=& \sum_{k=0}^{n_2-1}{}^{n_2-1}C_k(L_{-1} + J^{3}_{-1})^k(-1)^{n_2-1-k} 
\big((L_{-1} + J^{3}_{-1})\circ\big)^{n_2-1-k}d^{++(i)}_{-n_1}
\cr 
&=&\sum_{k=0}^{n_2-1}{}^{n_2-1}C_k(L_{-1} + J^{3}_{-1})^k(-1)^{n_2-1-k}{(n_1 + n_2-k-1)!\over n_1!}d^{++(i)}_{-(n_1 + n_2 -k-1 )} 
\eea
Inserting this into (\ref{C2}) gives
\bea
|\psi\rangle&=&{1\over (n_2-1)!}\sum_{k=0}^{n_2-1}{}^{n_2-1}C_k(-1)^{n_2-k-1}{(n_1 + n_2-k-1)!\over n_1!}\cr 
&&\hspace{2cm}\s^+_2(w)(L_{-1} + J^{3}_{-1})^k d^{++(i)}_{-(n_1 + n_2 -k-1 )}d^{--(j)}_{-1}|0^-_R\rangle^{(1)}|0^-_R\rangle^{(2)}
\eea
This can be written as 
\bea 
|\psi\rangle&=&{1\over n_1!(n_2-1)!}\sum_{k=0}^{n_2-1}{}^{n_2-1}C_k(-1)^{n_2-k-1}(n_1 + n_2-k-1)!\cr 
&&(L_{-1} + J^{3}_{-1} -e^{-w}\partial )^k\s^+_2(w) d^{++(i)}_{-(n_1 + n_2 -k-1 )}d^{--(j)}_{-1}|0^-_R\rangle^{(1)}|0^-_R\rangle^{(2)}\cr 
&=& {1\over n_1!(n_2-1)!}\sum_{k=0}^{n_2-1}{}^{n_2-1}C_k(-1)^{n_2-k-1}(n_1 + n_2-k-1)!\cr 
&&(L_{-1} + J^{3}_{-1} -e^{-w}\partial )^k\big( \sum_{p\geq 0}f^+_i[-(n_1+n_2-k-1),-p]d^{++}_{-p}\sum_{p'\geq0}f^-_j[-1,-p']d^{--}_{-p'}\cr 
&&+ C^{ij}_F[n_1+n_2-k-1,1] \big)|\chi\rangle
\eea 
Keeping only terms which are not proportional to any fermionic modes and comparing with (\ref{Cn1n2}) we obtain the relation
\bea\label{Cn1n2p}
C^{ij}_F[n_1,n_2] &=& {1\over n_1!(n_2-1)!}\sum_{k=0}^{n_2-1}{}^{n_2-1}C_k(-1)^{n_2-k-1}(n_1 + n_2-k-1)!\cr 
&&(-e^{-w}\partial)^kC^{ij}_F[n_1+n_2-k-1,1]
\eea
Using the expression in (\ref{Cn1}) we compute the following term
\bea
&&(-e^{-w}\partial)^kC^{ij}_F[n_1+n_2-k-1,1]\cr 
&&=C^{ij}_F[n_1+n_2-k-1,1]_{w=0}{(n_1 + n_2 - 1)!\over (n_1 + n_2 -k-1)!}e^{-(n_1+n_2)w}
\eea
Inserting this into (\ref{Cn1n2p}) yields
\bea
C^{ij}_F[n_1,n_2]= e^{-(n_1+n_2)w}{(n_1 + n_2 - 1)!\over n_1!(n_2-1)!}\sum_{k=0}^{n_2-1}{}^{n_2-1}C_k(-1)^{n_2-k-1}C^{ij}_F[n_1+n_2-k-1,1]_{w=0}\nn
\eea
Performing the sum yields
\bea 
    C^{ij}_F[n_1,n_2] &=&-C^2(-1)^{i+j}{2\Gamma[{1\over2}+n_1]\Gamma[{1\over2}+n_2]\over \pi n_1\Gamma[n_1]\Gamma[n_2]}{e^{-(n_1+n_2)w}\over n_1+n_2}
\eea

\bibliographystyle{JHEP}
\bibliography{bibliography.bib}

\end{document}